\documentstyle[psfig,12pt]{article}
\def\lsim{\:\raisebox{-0.5ex}{$\stackrel{\textstyle<}{\sim}$}\:}

\def\be{\begin{equation}}       
\def\ee{\end{equation}}
\def\bear{\be\begin{array}}      
\def\eear{\end{array}\ee}
\def\bea{\begin{eqnarray}}
\def\eea{\end{eqnarray}}

\include{mat}
\def\21{$SU(2) \ot U(1)$}
\def\ot{\otimes}
\def\ie{{\it i.e.}}

\def\half{{\textstyle{1 \over 2}}}

\def\quarter{{\textstyle{1 \over 4}}}

\def\eighth{{\textstyle{1 \over 8}}}
\def\bold#1{\setbox0=\hbox{$#1$}
     \kern-.025em\copy0\kern-\wd0
     \kern.05em\copy0\kern-\wd0
     \kern-.025em\raise.0433em\box0 }
\begin{document}
\begin{titlepage}
\begin{flushright}
IFIC/98-17\\
FTUV/98-17\\
hep-ph/9802275\\
February 1998
\end{flushright}
\vspace*{5mm}
\begin{center} 
{\Large \bf Charged Scalar Phenomenology in the Bilinear R--Parity
Breaking Model${}^{\dag}$}\\[15mm]
{\large Javier Ferrandis} \\
\hspace{3cm}\\
{\small Departamento de F\'\i sica Te\'orica, IFIC-CSIC, 
Universidad de Valencia}\\ 
{\small Burjassot, Valencia 46100, Spain}
\end{center}
\vspace{5mm}
\begin{abstract}

We consider the charged scalar boson phenomenology in the bilinear 
R-parity breaking model which induces a 
mixing between staus and the charged Higgs boson.The 
charged Higgs boson mass can be lower than expected in the MSSM, even 
before including radiative corrections.The R-parity 
violating decay rates can be comparable or even bigger than the 
R-parity conserving ones. These features could 
have implications for charged supersymmetric scalar boson 
searches at future accelerators.

\end{abstract}

\vskip 5.cm
\noindent ${}^{\dag}$Talk given at the International Workshop
``Beyond the Standard Model: From Theory to Experiment'', 
13--17 October 1997, Valencia, Spain.

\end{titlepage}

\setcounter{page}{1}

 A lot of emphasis has been put into the phenomenological study of the 
supersymmetric Higgs boson sector \cite{carena}. However, so far most 
of these phenomenological studies have been made in the framework of 
the Minimal Supersymmetric Standard Model with conserved 
R-parity \cite{RP}.  
Alternative supersymmetric scenarios where the effective 
low energy theory violates R-parity \cite{HallSuzuki} have recently 
received a lot of attention \cite{Dreiner,beyond}.
The simplest version of these models, in which the violation 
of R-parity is effectively parametrized by a bilinear term 
$\epsilon_i\widehat L_i\widehat H_u$ is receiving growing attention
\cite{rpheno,charhig}.
We focus on the phenomenology of the charged scalar bosonic sector.
This complements a previous study of the electrically neutral sector
\cite{eps0}.
We show that: 1) the mass of the charged Higgs boson can be lower than expected in the MSSM, even before including radiative corrections, 
2) if the stau is the LSP it will only have R-parity violating decay 
channels into standard model fermions, while in the opposite case when
it is heavier than the lightest neutralino one expects exotic high multiplicity
events, 3) the branching ratios for the R-parity violating charged Higgs boson 
decays can be comparable or even bigger than the R-parity conserving 
ones. 

The Lagrangian is specified by the superpotential $W$ given by 
$$
W=\varepsilon_{ab}\left[ 
 h_U^{ij}\widehat Q_i^a\widehat U_j\widehat H_u^b 
+h_D^{ij}\widehat Q_i^b\widehat D_j\widehat H_d^a 
+h_E^{ij}\widehat L_i^b\widehat R_j\widehat H_d^a 
-\mu\widehat H_d^a\widehat H_u^b 
+\epsilon_i\widehat L_i^a\widehat H_u^b\right] 
$$
Supersymmetry breaking is parametrized by the set of soft 
supersymmetry breaking terms which do not introduce quadratic 
divergences to the unrenormalized theory. 
We will focus for simplicity on the case of one generation, 
namely the third \cite{eps0,sensi}.   
In contrast we keep in our discussion the theory as 
defined at low energies by the most general set of soft-breaking 
masses, trilinear and bilinear  soft-breaking parameters, 
gaugino masses and the Higgs superfield mixing  parameter $\mu$. 
The electroweak symmetry is broken when the two Higgs doublets  
$H_d$ and $H_u$, and the third component of the left slepton 
doublet $\widetilde L_3$ acquire vacuum expectation values to minimize 
the scalar potential. 
Note that the $W$ boson acquires a mass 
$m_W^2=\quarter g^2(v_u^2+v_d^2+v_{\tau}^2)$. \ 
In addition to the MSSM parameters, our model contains three new 
parameters, $\epsilon_3$, $v_{\tau}$ and $B_2$, of which only two are 
independent, and these may be chosen as $\epsilon_3$ and $v_{\tau}$.  
An important feature of this model is that lepton number is also 
violated by the $\epsilon_3$ term and by the presence of the sneutrino 
vacuum expectation value $v_{\tau}$. This induces a mass for the tau 
neutrino, which turns out to be:
$$m_{\nu_{\tau}} \approx \frac{(g^2 M+g^{\prime 2} M^{\prime })( \mu v_{\tau}
+ \epsilon_3 v_d)^2}
 {-4MM^{\prime}\mu^2 + 2\mu v_d v_u (g^2 M+g^{\prime 2} M^{\prime })+ 
 \delta( \epsilon_3,v_{\tau})} $$ 
with $ \delta( \epsilon_3,v_{\tau})=[2\epsilon_3 v_ {\tau} 
(\mu v_d - M^{\prime }v_u)+ 
v_{\tau}^2 \mu^2 ](g^2+g^{\prime 2})+\epsilon_3^2(g^2v_u^2 +
g^{\prime 2}v_d^2-4MM^{\prime })$
and it is naturally small in models with 
universality of soft mass parameters at the unification scale 
\cite{epsrad}. 
If the bilinear terms are rotated away the rotation will generate
trilinear and bilinear R--parity violating couplings in the scalar
potential which are proportional to $\epsilon_3$.
It has been claimed \cite{mv90,RPeps} that $v_{\tau}$ terms are proportional to
neutrino masses but in this basis may be possible to find light neutrinos 
with large $\epsilon_3$ and $\nu_{\tau}$ because 
$m_{\nu_{\tau}} \sim (\mu v_{\tau} + \epsilon_3 v_d)^2$ and small neutrino 
masses requires that the combination $(\mu v_{\tau} + \epsilon_3 v_d)$ 
be small.
Then we consider the possiblity of large R--violation.
In this paper we take the parameters at the weak scale 
independent (\ie\, no universality assumption), and allways impose 
$m_{\nu_{\tau}} \lsim 20$ MeV. 
The charginos mix with the tau lepton forming a set of three charged fermions.
The tau Yukawa coupling $h_{\tau}$ can be fixed , such that one of the  
singular values is equal to the tau mass , through an exact tree level formula
\cite{charhig}.

The $4\times4$ mass matrix of the charged scalar sector has three 
components in the minimum of the potential: 
$
\bf{M_{S^{\pm}}^2} = \bf {M_{H}^2} +
\bf{M_{\tilde\tau}^2} +
\bf{M_{\epsilon}^2}
$
where $\bf M^2_{H}$ is the $2\times2$ MSSM charged Higgs block and
$\bf M^2_{\tilde\tau}$ is the $2\times2$ MSSM stau block. 	 
The component not present in the MSSM which produces a mixing between 
the charged Higgs sector and the stau sector,\ $\bf {M_{\epsilon}^2}$
\ , is given by :
$$
\small{
\!\!\!\!\!\!\left[\matrix{
\half h_{\tau}^2v_{\tau}^2-\eighth(g^2-g'^2)v_{\tau}^2 &
\!\!\!\!\!\!\!\!\!\!\!\!\!\!\! 0 & 
\!\!\!\!\!\!\!\!\!\!\!\!\!\!\!\! \quarter(g^2-2h_{\tau}^2)v_dv_{\tau}-\mu\epsilon_3 & 
\!\!\!-{1\over{\sqrt{2}}}h_{\tau}(\epsilon_3v_u+A_{\tau}v_{\tau}) \cr 
0 & 
\!\!\!\!\!\!\!\!\!\!\!\!\!\!\!\! \epsilon_3^2+\eighth(g^2-g'^2)v_{\tau}^2 & 
\!\!\!\! -B_2\epsilon_3+\quarter g^2v_uv_{\tau} & 
\!\!\!\! -{1\over{\sqrt{2}}}h_{\tau}(\mu v_{\tau}+\epsilon_3v_d) \cr
\quarter(g^2-2h_{\tau}^2)v_dv_{\tau}-\mu\epsilon_3 & 
\!\!\!\! -B_2\epsilon_3+\quarter g^2v_uv_{\tau} & 
\!\!\!\! \epsilon_3^2+\eighth(g^2+g'^2)v_{\tau}^2 & 0 \cr
\!\!\!\! -{1\over{\sqrt{2}}}h_{\tau}(\epsilon_3v_u+A_{\tau}v_{\tau}) & 
\!\!\!\! -{1\over{\sqrt{2}}}h_{\tau}(\mu v_{\tau}+\epsilon_3v_d) & 0 & 
\!\!\!\! \half h_{\tau}^2v_{\tau}^2-\quarter g'^2v_{\tau}^2
}\right]
}
$$
\begin{figure} 
\centerline{\protect\hbox{
\psfig{figure=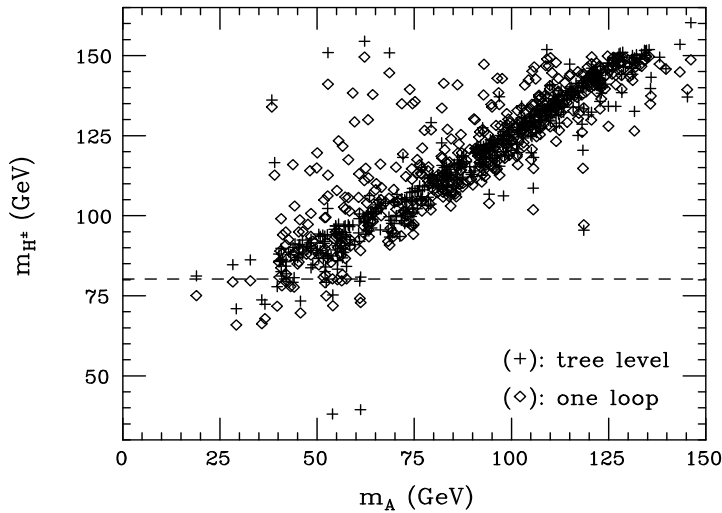,width=0.6\textwidth,height=7cm,angle=90}
}}
\caption{Tree level and one--loop charged Higgs boson mass  
as a function of the CP--odd Higgs mass $m_A$.} 
\label{fig:mchma} 
\end{figure}  
In our numerical study of the charged Higgs mass spectrum
, we have varied the MSSM parameters over a wide range.
Note that $m^2_{H_d}$, $m^2_{H_u}$, and $B_2$ are fixed through
minimization. No big differences are
observed if we change the sign of  $ \mu $.
We are interested in a relatively light charged Higgs , so we
take $ | \mu | $ , $ | B | \leq 200 $ \, GeV . Similarly we
are interested in relatively light staus , and that is why we take $
m_{R_{3}} $ , $ m_{L_{3}} $ $ \leq 300\: $ GeV.
The main point to note is that $m_{H^{\pm}}$ can be 
lower than expected in the MSSM, even before including radiative 
corrections. This is due to negative contributions arising from the 
R-parity violating stau-Higgs mixing, controlled by the parameter 
$\epsilon_3$. 

\begin{figure}
 \centerline{\protect\hbox{
\psfig{figure=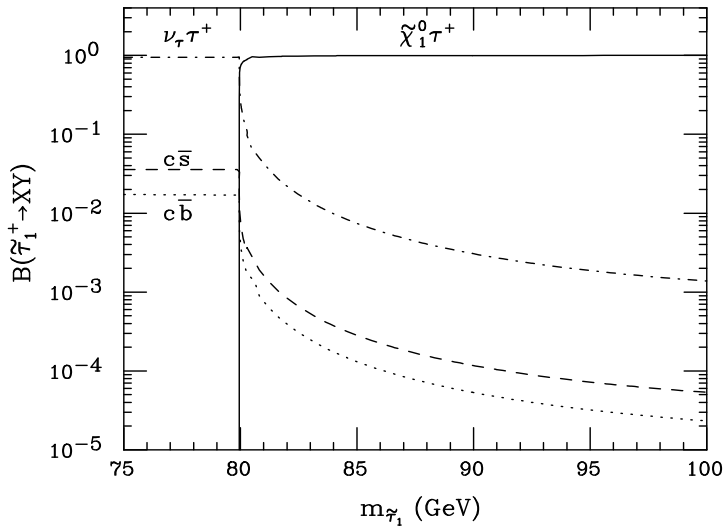,width=0.6\textwidth,height=7cm,angle=90}
}}
\caption{Stau branching ratios possible in our model. Note the neutralino threshold below which only R--parity violating decays are present.}  
\label{fig:v3stB} 
\end{figure}

We now turn to a discussion of the charged scalar boson decays. 
In Fig.~\ref{fig:v3stB} we display 
the stau decay branching ratios below and above the neutralino 
threshold. We have fixed the parameters 
in such a way as to ensure that the lightest neutralino is about 
80 GeV in mass and thus may be produced as a decay product of a stau 
produced at LEP II energies. Below the neutralino threshold the stau is 
the LSP and will have only R-parity violating decays :
$\nu_{\tau}\tau$, $c\bar{s}$ and $c\bar{b}$. 

\begin{figure}
 \centerline{\protect\hbox{
\psfig{figure=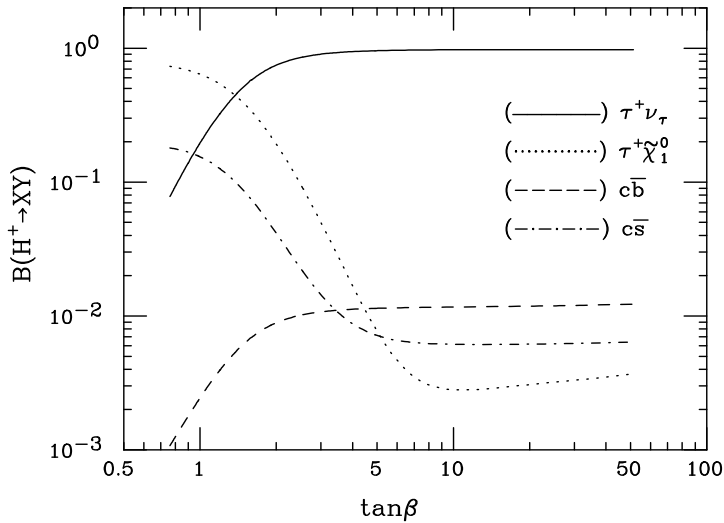,width=0.6\textwidth,height=7cm,angle=90}
}}
\caption{Charged Higgs branching ratios possible in our model for a
particular choice of parameters. The R--parity violating decay
dominates at low $\tan\beta$.}
\label{fig:h+br} 
\end{figure}

The charged Higgs boson branching ratios ,Fig.~\ref{fig:h+br}, into 
supersymmetric channels can be comparable or even bigger than the 
R-parity conserving ones, even for relatively small values of 
$\epsilon$ and $v_3$. 
In the region of small $\tan\beta$, the R-parity violating Higgs boson decay 
branching ratios can exceed the conventional ones and may reach values 
close to 1.
\noindent
We have scanned the parameter space and we find that even for very 
small R--parity violating parameters the branching ratio 
$B(H^{+} \longrightarrow \tau^{+} \tilde\chi^{0}_{1}) $ 
can be close to unity, and that in the
region of $\tan\beta \gg 1 $ the decay $ H^{+} \longrightarrow
\tau^{+} \nu_{\tau} $ dominates.

If the stau is the LSP it will decay only through 
R-parity-violating interactions, to cs or $\tau\nu_{\tau}$. Then 
it leads to signatures identical to those of 
the charged Higgs boson in the MSSM. 
However, if it is not the LSP the 
$\tilde\tau_1^{\pm}$ is more likely to have standard 
R-parity-conserving decays such as neutralino plus $\tau$, leading to 
signals that can be drastically different from those expected in the 
MSSM and which would arise from $\tilde\chi^0\longrightarrow 
\nu_{\tau}Z^*$ or $\tilde\chi^0\longrightarrow \tau W^*$. Unless 
$\epsilon_3$ and $v_3$ are extremely small, the neutralino will decay 
inside the detector. For the case of stau pair production in $e^+e^-$ 
colliders, such as LEP II, this would imply a plethora of new high 
fermion-multiplicity events (multi-jets and/or multi-leptons) . 
For example, 2tau + 4j + missing energy if both neutralinos decay 
into jets through neutral currents, or 4taus + 4j if both 
neutralinos decay into jets through charged currents. As for 
hadron colliders, we can also have very high leptonic multiplicity 
events such as six leptons of which at least two are taus, plus 
missing momentum. This should be easy to see at the LHC, due again to 
the negligible standard model background. 

Another interesting feature of our model is the mixed production 
$e^+e^-\longrightarrow H^{\pm}\tilde\tau^{\mp}$ which is absent in the MSSM.
If $m_{\chi^0_1}<m_{\tilde\tau_1^{\pm}}$ then one can produce interesting 
signatures like di-tau + di-jets + missing energy.  This is obtained 
when $H^{\pm}\longrightarrow \tau^{\pm}\nu_{\tau}$ and 
$\tilde\tau^{\mp}\longrightarrow\tau^{\mp}\tilde\chi^0_1 
\longrightarrow \tau^{\mp}q\overline{q}\nu_{\tau}$, and may have a 
non-negligible cross-section. 

\section*{Acknowledgments:} 
This work was supported by DGICYT under grants PB95-1077 and by the 
TMR network grant ERBFMRXCT960090 of the European Union.  
The author is indebted to his collaborators A. Akeroyd, 
M.A. D\'\i az, M.A. Garcia--Jare\~no, J.C. Rom\~ao, 
and J.W.F. Valle.
The author was supported by a Spanish MEC FPI fellowship.


\begin{thebibliography}{99} 
 
\bibitem{carena}  
Carena M.,Zerwas P., et. al.,[hep-ph 9602250]
 
\bibitem{RP} 
Farrar G.,Fayet P.,{\sl Phys.Lett.}{\bf 76B},575(1978); 
{\sl Phys. Lett.}{\bf 79B},442(1978). 
 
\bibitem{HallSuzuki} 
Hall L. and Suzuki M., {\sl Nucl.Phys.} {\bf B231},419(1984).
 
\bibitem{Dreiner} 
Dreiner H. [hep-ph/9707435]

\bibitem{beyond} 
Valle J. W.[hep-ph/9603307]. 
 
\bibitem{rpheno}
Carena M.,Pokorski S.,Wagner C.E.M. [hep-ph/9801251];
Faessler A.,Kovalenko S., Simkovic F. [hep-ph/9712535];
D\'\i az M.,Rom\~ao J.,Valle J. W., [hep-ph/9706315];
Hempfling R.,{\sl Nucl.Phys.B478:3-30,1996};
Roy S.,Mukhopadhyaya B.{\sl Phys.Rev.D55:7020-70};
Davidson S.,Hempfling R. {\sl Phys.Lett.B391:287-294,1997} 

\bibitem{charhig}
Akeroyd A.,D\'\i az M.,Ferrandis J.,Garc\'\i a Jare\~no M.,
Valle J.W ,[hep-ph 9707395].

\bibitem{eps0}
de Campos F., Garc{\'\i}a-Jare\~no M.,Joshipura A.,Rosiek J., 
Valle J.,{\sl Nucl. Phys.}{\bf B451},3(1995);

\bibitem{sensi}  
Romao J.,de Campos F.,Garc\'\i a-Jare\~no M.,Magro M., 
Valle J. W.,Nucl.Phys.B482(1996)3-23 
 
\bibitem{mv90} 
Masiero A.,Valle J. W.,{\sl Phys.Lett.}{\bf B251},273(1990); 
for other ref. see \cite{beyond}. 
 
\bibitem{RPeps}  
Vissani F.,Smirnov A.,{\sl Nucl. Phys.} {\bf B460}, 37 (1996);  
Nilles H. P.,Polonsky N., {\sl Nucl. Phys.} {\bf B484}, 33 (1997);  
de Carlos B.,White P. L., {\sl Phys. Rev.} {\bf D55}, 4222 (1997);  
   

\end{thebibliography}
\end{document}